\documentclass[preprint,eqsecnum,preprintnumbers,nofootinbib,prd,aps,showpacs,showkeys,groupedaddress,
floatfix]{revtex4-1}
\usepackage{bm}
\usepackage{graphics}
\usepackage{graphicx}
\usepackage{epsfig}
\usepackage{amssymb}
\usepackage{amsmath}

\newcommand\ba{\begin{eqnarray}}
\newcommand\ea{\end{eqnarray}}

\begin{document}
%\large

\title{Spectra of Heavy Quarkonia in a Magnetized-Hot Medium in the Framework of Fractional Non-relativistic Quark Model}
\author{M. Abu-shady$^{1}$ \footnote{E-mail: dr.abushady@gmail.com}}
\author{A. I. Ahmadov$^{2,3}$ \footnote{E-mail: ahmadovazar@yahoo.com}}
\author{H. M. Fath-Allah$^{4}$ \footnote{E-mail: hebamosad8@gmail.com}}
\author{V. H. Badalov$^{3}$ \footnote{E-mail: badalovvatan@yahoo.com}}

\affiliation {$^{1}$ Department of Mathematics and Computer Sciences, Faculty of Science, Menoufia University, Menoufia, Egypt\\
$^{2}$Department of Theoretical Physics, Baku State University, Z. Khalilov St. 23, AZ-1148 Baku, Azerbaijan\\
$^{3}$Institute  for Physical Problems, Baku State University,\\ Z. Khalilov st. 23, AZ-1148, Baku, Azerbaijan\\
$^{4}$ Higher Institute of Engineering and Technology, Menoufia, Egypt}

\date{\today}

\begin{abstract}
In the fractional nonrelativistic potential model, the decomposition
of heavy quarkonium in a hot magnetized medium is investigated. The
analytical solution of the fractional radial Schr\"{o}dinger
equation for the hot-magnetized interaction potential is displayed
by using the conformable fractional Nikiforov-Uvarov method.
Analytical expressions for the energy eigenvalues and the radial
wave function are obtained for arbitrary quantum numbers. Next, we
study the charmonium and bottmonium binding energies for different
magnetic field values in the thermal medium. The effect of the
fractional parameter on the decomposition temperature is also
analyzed for charmonium and bottomonium in the presence of hot
magnetized media. We conclude that the dissociation of heavy
quarkonium in the fractional nonrelativistic potential model is more
practical than the classical nonrelativistic potential model.

\end{abstract}
\maketitle

\section{INTRODUCTION}\label{1}

Quantum chromodynamics theory calculates that at sufficiently high
temperatures and densities, the gluons and quarks confined inside
the hadrons are freed into a medium of gluons and quarks. Recent
works have focused on producing and identifying this new state of
matter theoretically [1-7] and experimentally in ultra-relativistic
heavy-ion collisions (URHIC) with the increasing center of mass
energies in the BNL AGS, CERN SPS, BNL RHIC, and CERN LHC
experiments. However, for the noncentral events in URHICs, the
powerful magnetic field is generated at the collisions' initial
stages due to very high relative velocities of the spectator quarks
concerning the fireball [8,9].

There are numerous research studies to investigate the properties of
quarkonium in the magnetic field [10], in which the
three-dimensional Schr\"{o}dinger equation (SE) numerically solved
with the Cornell and the QCD Coulomb potentials. In Ref. [11], the
properties of quarkonium states have been studied in the presence of
strong magnetic field. Two methods were used to calculate the
critical value of the magnetic field for both charmonium and
bottomonium states. Bagchi \textit{et. al.} inferred that in the
presence of a magnetic field, the bound states $J/\psi$ and
$Y(1S)$become more firmly bound than in a pure thermal QGP owing to
the alteration of the heavy quark potential [12]. In Ref. [13], the
authors studied the effect of a strong external magnetic field on
quarkonium states $c\bar{c}$ and $b\bar{b}$ in the framework of a
non-relativistic quark model. Furthermore, the authors included in
their calculation anisotropies through static quark-antiquark
potential in agreement with recent lattice studies. In Ref. [14],
the dissociation of heavy quarks in hot QCD plasma in the presence
of a strong magnetic field is studied by using Nikiforov-Uvarov (NU)
method. In Ref. [15], one can get more recent review about heavy
quarkonium in extreme condition.

In recent years, there has been considerable interest in fractional
calculus of several branches of physics [16-21]. The
analytical-exact iteration method was extended to the conformable
fractional form to obtain the analytical solutions of the
N-dimensional radial SE with its applications on heavy mesons [16].
The generalized NU method was also extended to the fractional domain
of high-energy physics by using the radial SE [17]. Abdeljawad used
the fractional concept of NU to solve fractional radial SE for
different interaction potentials such as the oscillator potential,
Woods-Saxon potential, and Hulth\'en potential [18]. Furthermore,
Herrmann applied the derivative Caputo fractional Schr\"{o}dinger
wave equation by using quantitative of the classical nonrelativistic
Hamiltonian [19]. The conformable fractional form was extended to a
finite temperature medium to study the binding energy and
dissociation of temperature [20].

\noindent The aim of the present work is to find the binding energy
and the dissociation temperature of quarkonia by using fractional
non-relativistic quark model. It shows that the fractional model is
more practical to study like that problems. In addition, heavy
quarkonium has been thoroughly studied in a hot-magnetized medium.
This paper is organized as follows: In Sec. \ref{2}, we provide the
theoretical method. In Sec. \ref{3}, the method is given in detail
to solve the N-dimensional SE. In Sec. \ref{4}, we discuss the
obtained results. The conclusion is given in Sec. \ref{5}.

\section{Theoretical model}\label{2}

Fractional derivative plays an important role in the applied
science. Riemann-Liouville and Riesz and Caputo gave an elegant
formula that allows to apply boundary and initial conditions as in
Ref. [21].
\begin{equation} \label{GrindEQ__1_}
{D_t}^{\alpha }\left(r\right)=\int^r_{r_0}{K_a(r-s)f^{\left(n\right)}(s)d(s)}, {r>r}_{0}
\end{equation}
with
\begin{equation} \label{GrindEQ__2_}
K_a\left(r-s\right)=\frac{{(r-s)}^{n-\alpha-1}{\rm \ }}{\Gamma(n-\alpha )}  ,
\end{equation}
where,  $f^{(n)}$ is the $n$ the derivative of the function f(r),
and $K_a(r-s)$ is the kernel, which is fixed for a given real number
$\alpha$. The kernel${\ K}_a(r-s)$ has singularity at $r=s$. Caputo
and Fabrizio[22] suggested a new formula of the fractional
derivative with smooth exponential kernel of the form to avoid the
difficulties that found in Eq. \eqref{GrindEQ__1_}

\begin{equation} \label{GrindEQ__3_}
{D_r}^{\alpha}=\frac{M\left(a\right)}{1-\alpha }\int^r_{r_0}{{\exp  \left(\frac{\alpha \left(t-s\right)}{1-\alpha }\right)\ }}\dot{y}\left(s\right)d(s),
\end{equation}
where $M(a)$ is a normalization function with
$M(0)=M\left(1\right)=1$.

A new formula of fractional derivative called conformable fractional derivative (CFD) was proposed by Khalil et al [23].
\begin{equation} \label{GrindEQ__4_}
{{D_t}^{\alpha }f\left(r\right)=\lim }_{\varepsilon \to 0} \frac{f\left(r-\varepsilon r^{1-\alpha }\right)-f(r)}{\in }\, , r>0
\end{equation}
\begin{equation} \label{GrindEQ__5_}
{\rm f}\left(0\right){\rm =lim}_{\varepsilon \to 0}\ f(r),
\end{equation}
where
\begin{equation} \label{GrindEQ__6_}
D^{\alpha }[\ f_{nl}(r{)]=r}^{1-\alpha }{\grave{f}}_{nl}(r),
\end{equation}
\begin{equation}  \label{GrindEQ__7_}
{D^{\alpha }[{\rm \ }D}^{\alpha }\ f(r)] = (1- \alpha) r^{1-2\ \alpha }{\grave{f}}_{nl}(r) + r^{2-2{\alpha}}{{\ f}_{nl}}^{''}(r),                     
\end{equation}
with ${\rm 0\ }<\ \alpha {\rm \ }\leq{\rm \ \ 1}$. This a new
definition is simple and provides a natural extension of
differentiation with integer order  $n\in Z$ to fractional order
$\alpha \in C$. Moreover, the CFD operator is linear and satisfies
the interesting properties that traditional fractional derivatives
do not, such as the formula of the derivative of the product or
quotient of two functions and the chain rule [24].

\section{The solution of the radial Schr\"{o}dinger equation in the presence of a strong magnetic field}\label{3}
\noindent As in Refs. [16, 25], in the $N$-dimensional space, the SE
for two particles which interact with symmetrical potentials takes
following form
\begin{equation} \label{GrindEQ__8_}
\left[\frac{d^2}{dr^2}+\frac{N-1}{r}\frac{d}{dr}-\frac{l\left(l+N-2\right)}{r^2}+2\mu
(E-V\left(r\right))\right]\Psi(r)=0,
\end{equation}
where $l$, $N$ and $\mu$ are the angular momentum quantum number,
the dimensional number, and the reduced mass of the system. The
following radial SE is obtained by applying the wave function $\Psi
(r)=r^{\frac{1-N}{2}}R(r)$,
\begin{equation} \label{GrindEQ__9_}
\frac{d^2R(r)}{dr^2}-2\mu \left( E- V(r)-
\frac{{\left(l+\frac{N-2}{2}\right)}^2-\frac{1}{4}}{2\mu
r^2}\right)R(r)=0.
\end{equation}
The potential takes the form as in Ref.[25]:
\begin{equation} \label{GrindEQ__10_}
V\left(r\right)=-\frac{4}{3}\alpha \left({\rm}\frac{e^{-m_{D}{r}}}{r}+m_D\right)+\frac{4}{3}\frac{\sigma }{m_D}({\rm 1 -} e^{-m_{D}{r}})
\end{equation}
where, the string tension $\sigma $ = 0.18 Ge$V^2\ $and
\begin{equation} \label{GrindEQ__11_}
\alpha =\frac{12{\pi}}{11N_{c}{\ln (\frac{{{\mu}_0}^2+{M_B}^2}{{{\Lambda}_V}^2})}}
\end{equation}
where $N_c$ is the number of colors, $M_B$ ($\sim \ 1\ $GeV) is an
infrared mass interpreted as the ground state mass of the two gluons
bound to by the basic string, ${\ \mu }_0\ $= 1.1 (GeV), ${\Lambda
}_V$ = 0.385 (GeV) as in Refs. [27-29] and the Debye mass [29]
becomes as:
\begin{equation} \label{GrindEQ__12_}
{m_D}^2={g^2T}^2{\rm+}\frac{g^2}{4{\pi}^2{T}}{\rm}\sum_f{\left|q_fB\right|}{\rm}\int^{\infty}_0{\frac{e^{{\beta}\sqrt{{p_z}^2+{m_f}^2}}}{{(1+e^{{\beta}\sqrt{{p_z}^2+{m_f}^2}})}^{2}}}{\rm{d}}p_{z},
\end{equation}
where the first term is the contribution from the gluon loops and
dependent on temperature and the magnetic field does not affect it.
The second term is this term strongly depends on the magnetic field
$eB$ and is not much sensitive to the temperature $T$ of the medium.
In the first term, where $\acute{g}$ is the running strong coupling
constant and is given by

\begin{equation} \label{GrindEQ__13_}
\acute{g}=4{\pi}\acute{{\alpha}_s(T)},
\end{equation}
where, $\acute{{\alpha}_s}$(T) is the usual temperature-dependent
running coupling constant. It is given by
\begin{equation} \label{GrindEQ__14_}
\acute{{\alpha}_{s}}\left({\rm{T}}\right){\rm{=}}\frac{2{\pi}}{\left(11-\frac{2}{3}N_{f}\right){\ln  (\frac{\Lambda }{{\Lambda}_{QCD}})}} ,
\end{equation}
where, $N_f$ is the number of flavors, $\Lambda$ is the renormalization scale is taken as  $2\pi T$  and ${\Lambda }_{QCD}\sim $ 0.2 (GeV) as in Ref. [30].

The second term is $g$ =3.3, $q_f$ is the quark flavor $f=u$ and
$d$, $B$ is the magnetic field, $\beta$ is the inverse of
temperature and quark mass massive $m_{f}$ = 0.307 (GeV) as in Ref.
[31]. In Eq. \eqref{GrindEQ__12_}, $e^{-\ m_D\ r}$ is extend if
$m_Dr\ll$ 1 is considered in Ref.[32]. We rewrite Eq.
\eqref{GrindEQ__10_} as follows:

\begin{equation} \label{GrindEQ__15_}
{V\left(r\right)=a}_{1}r^{2}+a_{2}r+\frac{a_{3}}{r}
\end{equation}
where,
\begin{equation} \label{GrindEQ__16_}
a_{1}\ = -\frac{2}{3}\sigma{m_{D}},
\end{equation}
\begin{equation} \label{GrindEQ__17_}
a_{2} = -\frac{4}{3}{\alpha}{m_{D}}^{2}+\frac{4}{3}{\sigma},
\end{equation}
\begin{equation} \label{GrindEQ__18_}
a_{3} = -\frac{4}{3}{\alpha} .
\end{equation}
The fractional of Eq. \eqref{GrindEQ__8_} takes the following form as in Ref. [24]

\begin{equation} \label{GrindEQ__19_}
D^{\alpha}[D^{\alpha}{\Psi}\left(r^{\alpha}\right){{\rm}]}+\left[2{\mu}\left(E-V\left(r^{\alpha}\right)\right)-\frac{{\left(l+\frac{N-2}{2}\right)}^{2}-\frac{1}{4}}{2{\mu}{r}^{2\alpha}}\right]\Psi \left(r^{\alpha}\right){{\rm}={0}}
\end{equation}
the interaction potential in Eq. \eqref{GrindEQ__15_} is written in
the fractional form as in Refs. [16,17]
\begin{equation} \label{GrindEQ__20_}
{\rm{V}}\left(r^{\alpha}\right){{\rm}{=}}a_{1}{r}^{2{\alpha}}{{\rm}+}a_{2}{\rm}{r}^{\alpha}{{\rm}+}\frac{a_{3}}{r^{\alpha}}{\rm}.
\end{equation}
By applying NU method (For detail, see Refs. [17,18,20]), we obtain
the spectrum of energy

\begin{equation} \label{GrindEQ__21_}
E=\frac{{6}{a_{1}}}{{\delta}^2}+\frac{{3{a}}_{2}}{\delta}-\frac{2{\mu}{\left(\frac{8a_{1}}{{\delta}^{3}}{\rm +}\frac{3a_{2}}{{\delta}^{2}}-a_{3}\right)}^{2}}{{\left[\left(2n+1\right)\pm \sqrt{w+8{\mu} \left(\frac{3a_{1}}{{\delta}^{4}}\ {\rm}\frac{a_{2}}{{\delta}^{3}}{\rm+}\frac{{\left(l+\frac{N-2}{2}\right)}^{2}-\frac{1}{4}}{2{\mu}}\right)}\right]}^{2}},
\end{equation}
where
\begin{equation} \label{GrindEQ__22_}
{w=(2n{\alpha})}^{2}-4(n\left(3{\alpha}-{\alpha}^{2}\right)+\frac{1}{2}n(n-1){\alpha}\left({\alpha} +1\right)+{\alpha}-1)
\end{equation}
The radial wave function takes the following form:
\begin{equation} \label{GrindEQ__23_}
R_{nl}(r^{\alpha})=C_{nl}r^{(-\frac{B_{1}}{\sqrt{2A_{1}}}-1){\alpha}}{e}^{\sqrt{2A_{1}}r^{\alpha}}(-r^{2\alpha}D)^{n}(r^{(-2n+\frac{B_{1}}{\sqrt{2A_{1}}})\alpha}e^{-2\sqrt{2A_{1}}{r}^{2{\alpha}}})
\end{equation}
$C_{nL}$ is the normalization constant, and
\begin{equation} \label{GrindEQ__24_}
A_{1}=-{\mu}(E-\frac{6a_{1}}{{\delta}^{2}}-\frac{3{a}_2}{\delta}),
\end{equation}
\begin{equation} \label{GrindEQ__25_}
B_{1}={\mu}(\frac{8a_{1}}{{\delta}^{3}}+\frac{3a_{2}}{{\delta}^{2}}-a_{3})
\end{equation}
\begin{equation} \label{GrindEQ__26_}
C_{1}={\mu}(\frac{3a_{1}}{{\delta}^{4}}+\frac{a_{2}}{{\delta}^{3}}+\frac{{(l+\frac{N-2}{2})}^{2}-\frac{1}{4}}{2{\mu}}).
\end{equation}
\section{Results and Discussion}\label{4}
\begin{figure}
\includegraphics[width=\textwidth]{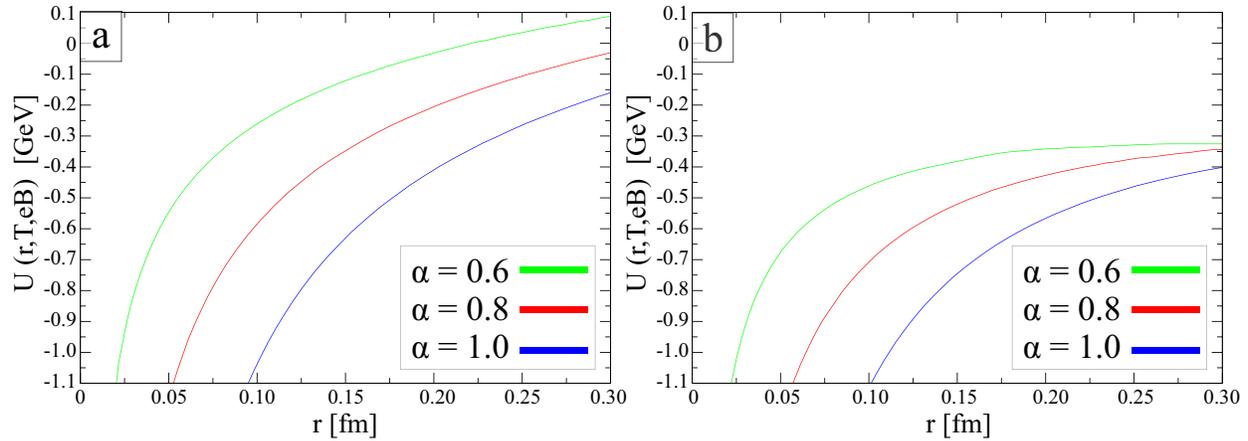}
\caption{\label{fig-1}(Color online) The potential interaction as a function of (r) with dependent on different parameters ${\alpha}$ at (a) $T=T_{c}$ and (b) $T=2T_{c}$.}
\end{figure}

\begin{figure}
\includegraphics[width=10 cm]{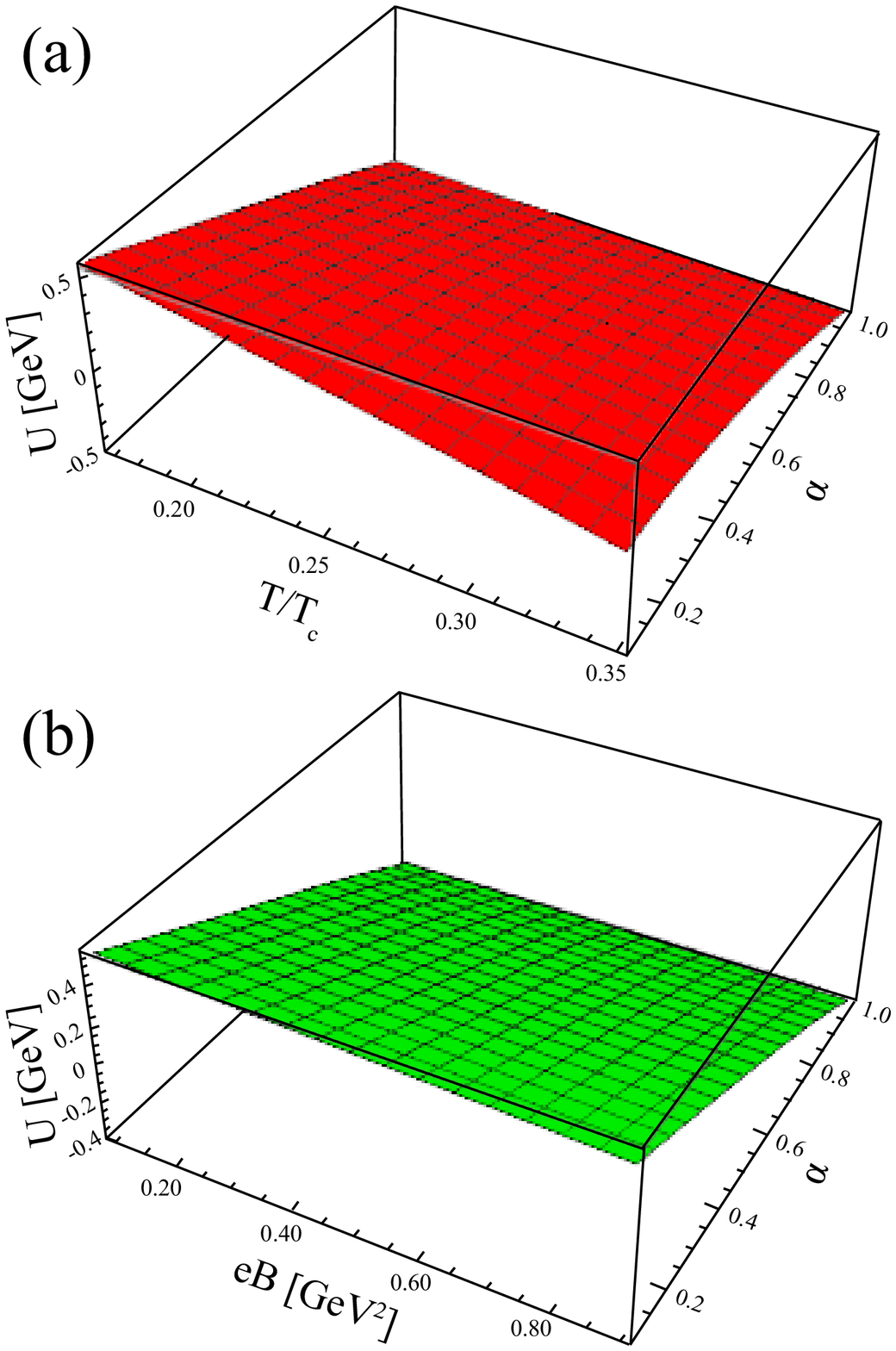}
\caption{\label{fig-2}(Color online) The potential interaction as (a) a function of temperature ratio and (b) magnetic field with dependent on fractional parameter.}
\end{figure}

The Fig. \ref{fig-1} indicates that the real-part is more screened by increasing fractional parameter. Besides, the fractional parameter has an effect on the linear term of potential. In addition, the potential becomes more attractive by increasing temperature from ${T}_{c}$ to ${2T}_{c}$.
In Fig. \ref{fig-2} (a), the real part of potential is plotted as a
function of the temperature ratio and magnetic field with dependent
on the fractional parameters for the fixed value of $r=0.2$ fm. By
taking the temperature range $T=0.17-0.3$ GeV, the potential is more
attractive with increasing fractional parameter than temperature.

In Fig. \ref{fig-2} (b), we see that the potential interaction is
more screened by increasing magnetic field and fractional parameter.
Thus, we deduce that the fractional parameter plays a role in the
hot medium at fixed magnetic field and the magnetized medium at
fixed temperature.

The QGP is distinguished by color screening: the range of
interaction between heavy quarks becomes inversely proportional to
temperature. At sufficiently high temperatures, forming a bound
state with a heavy quark ($c$ or $b$) and its anti-quark becomes
impossible. With increasing temperature, the range of interaction
decreases. For temperatures above the transition temperature, $T_c$,
the heavy quark interaction range becomes comparable to the
charmonium radius. Based on this general observation, one would expect that the charmonium states, as well as the  bottomonium states, do not exist above the
deconfinement transition.

\subsection{Binding energy}
\begin{figure}
\includegraphics[width=\textwidth]{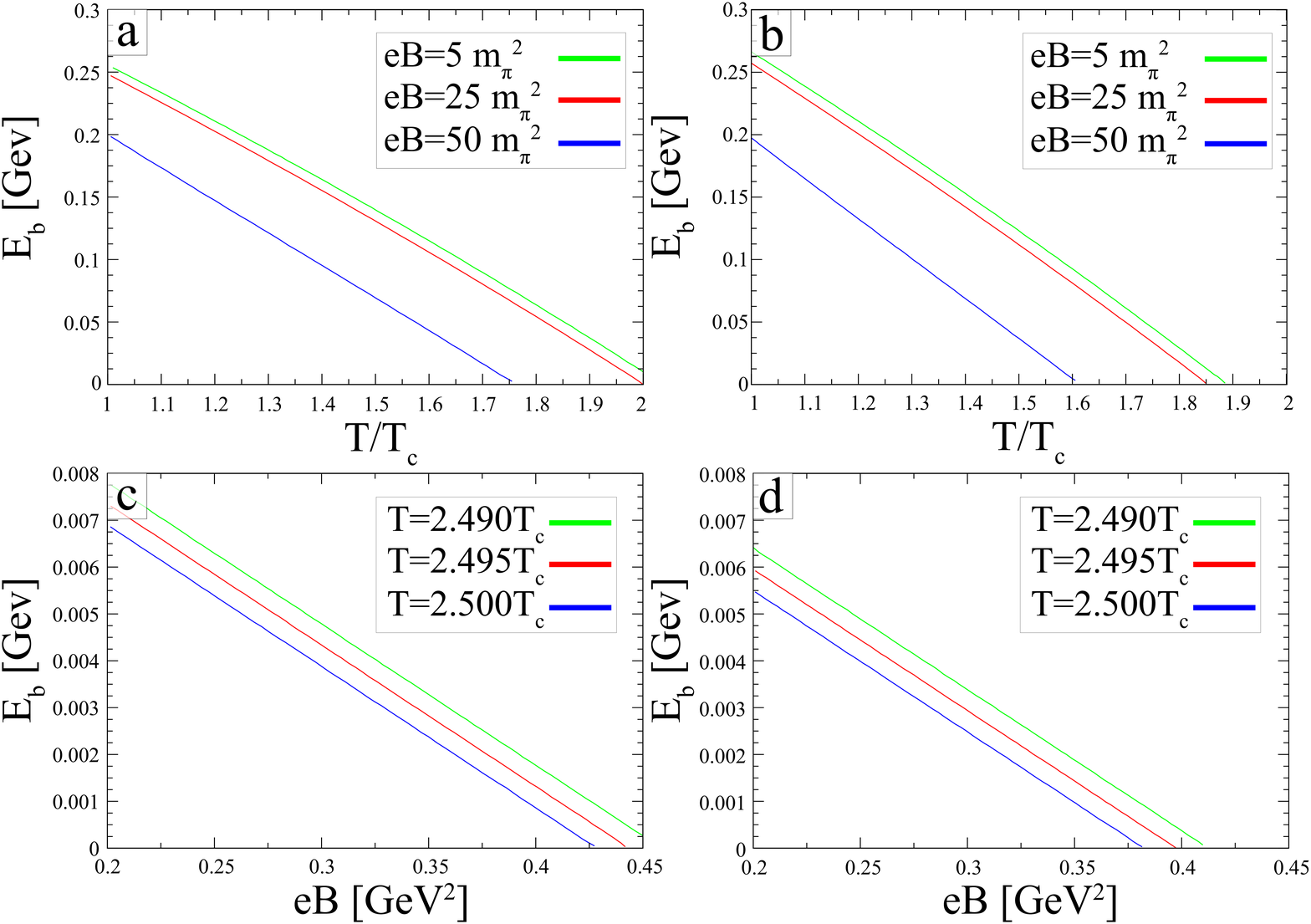}
\caption{\label{fig-3}(Color online) The binding energy $E_{b}$ of
charmonium as a function of the $T$ in the thermal medium with
dependent on the different $eB$ magnetic field values at (a)
$N_{f}=2$; ${\alpha}=1$ and (b) $N_{f}=2$; ${\alpha}=0.5$. The
binding energy $E_{b}$ of charmonium as a function of the magnetic
field $eB$ in the thermal medium with dependent on different values
of the temperature at (c) $N_{f}=2$; ${\alpha}=1$ and (d) $N_{f}=2$;
${\alpha}=0.5$.}
\end{figure}

\begin{figure}
\includegraphics[width=\textwidth]{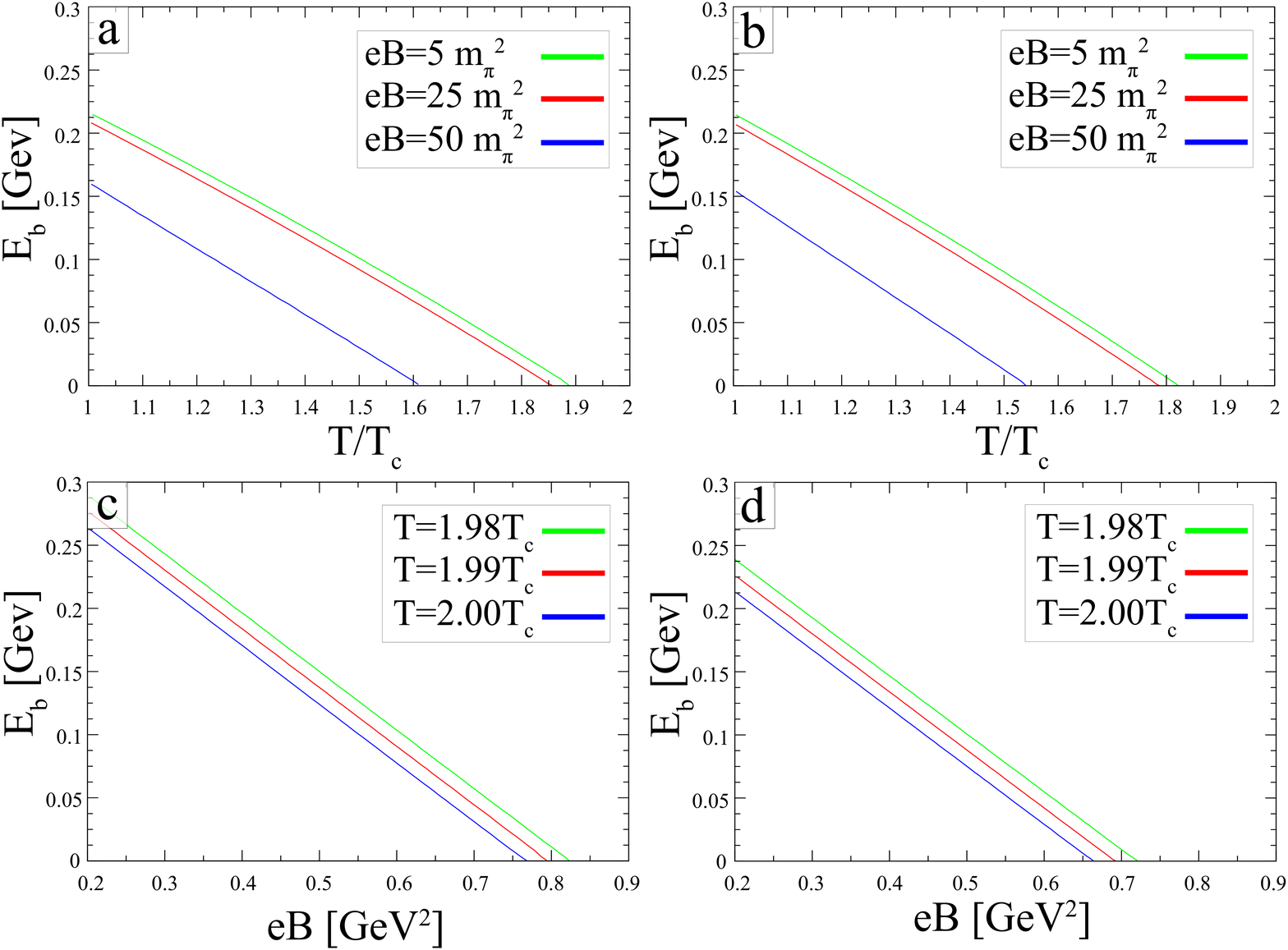}
\caption{\label{fig-4}(Color online) The binding energy $E_{b}$ of bottomonium as a function of the $T$ in the thermal medium with dependent on the different $eB$ magnetic field values at (a) $N_{f}=2$; ${\alpha}=1$ and (b) $N_{f}=2$; ${\alpha}=0.5$. The binding energy $E_{b}$ of bottomonium as a function of the magnetic field $eB$ in the thermal medium with dependent on different values of the temperature at (c) $N_{f}=2$; ${\alpha}=1$ and (d) $N_{f}=2$; ${\alpha}=0.5$.}
\end{figure}

By solving the radial SE, we obtain the binding energies $E_{b}$ of
$c\bar{c}$ and $\ b\bar{b}$. In following, we see the change of the
binding energy under the effect of fractional parameter in the
hot-magnetized medium. Charmonium binding energy $E_{b}$ is plotted
in Fig. \ref{fig-3}. The binding energy of charmonium decreases with
increasing temperature as well as the binding energy shifts to lower
values by increasing magnetic field at fixed $\alpha {\rm =1}$.(See
Fig \ref{fig-3}(a)) The binding energy go to zero energy faster by
decreasing fractional parameter $\alpha$. It indicates that the
dissociation of temperature are affected with depeding on fractional
parameter.
Similarly, the binding energy decreases when the magnetic field
increases. (See the Fig.\ref{fig-3}(c,d)) The binding energy shifts
slightly to the lower values by increasing temperature of medium.
Hence, the binding temperature tends to zero when temperature
increases. Furthermore, the binding energy is faster to tends to
zero by decreasing fractional parameter from $\alpha {\rm =1}$ to
$\alpha {\rm =0.5}$. As seen from Fig. \ref{fig-4}, the binding
energy of $1S$ bottomonium decreases by increasing temperature.
Besides, the binding energy decreases by increasing magnetic field.
While comparing Fig. \ref{fig-4}(a,b), it is seen that the
binding energy tends to zero rapidly by decreasing fractional
parameter from $\alpha{\rm =1}$ to $\alpha{\rm =0.5}$. Furthermore,
while comparing Fig. \ref{fig-4}(c,d), it is seen that the binding
energy $E_{b}$ decreases with increasing temperature $T$ and
magnetic field $eB$. The binding energy tends to zero
faster the left panel by decreasing fractional parameter from $\alpha{\rm =1}$ to $\alpha{\rm =0.5}$, too.

\subsection{Dissociation temperature with fractional parameter}
In the present work, we obtain the dissociation temperature at $E_{b}\simeq 0$, the approximation provides good accuracy in calculating the dissociation temperature. In the current analysis, we also study influence of the fractional parameter on the dissociation temperature in the presence of hot-magnetized medium for charmonium and bottomonium, using the calculated binding energies.

Table \ref{table1} presents the effect fractional parameter on the dissociation of temperature for different values of magnetic field. It is seen that the dissociation of temperature decreases by increasing magnetic field at ${\alpha}=1$. By taking ${\alpha}=0.5$, the dissociation of temperature gets lower values than the values at ${\alpha}=1$. Similarly, the dissociation of bottomonum decreases by decreasing fractional parameter. As a result, the dissociation of bottomonium is lower than the dissociation of charmonium.
\begin{table}
\centering
\caption{Dissociation temperature ($T_{D}$) for charmonium}
\label{table1}
\begin{tabular}{|p{1.0in}|p{1.1in}|p{1.1in}|p{1.1in}|} \hline
State          & $eB=5{m_{\pi}}^2$ & $eB=25{m_{\pi}}^2$ & $eB=50{m_{\pi}}^2$ \\ \hline
${\alpha}=1$   & 2.50 $T_{c}$ & 2.00 $T_{c}$ & 1.72 $T_{c}$ \\ \hline
${\alpha}=0.5$ & 1.91 $T_{c}$ & 1.86 $T_{c}$ & 1.69 $T_{c}$ \\ \hline
\end{tabular}
\end{table}
\begin{table}
\centering
\caption{Dissociation temperature ($T_{D}$) for bottomonium}
\label{table2}
\begin{tabular}{|p{1.0in}|p{1.1in}|p{1.1in}|p{1.1in}|} \hline
State          &$eB=5 {m_{\pi}}^2$ &$eB=25{m_{\pi}}^2$ & $eB=50{m_{\pi}}^2$ \\ \hline
${\alpha}=1$   &1.99$T_{c}$ & 1.85$T_{c}$ & 1.61$T_{c}$ \\ \hline
${\alpha}=0.5$ &1.82$T_{c}$ & 1.78$T_{c}$ & 1.55$T_{c}$ \\ \hline
\end{tabular}
\end{table}

\subsection{Dissociation of heavy quarkonia in a magnetic field}
We calculate the dissociation of charmonium and bottomonium at fixed temperature and fractional parameter when $E_{b}\simeq{0}$.
\begin{table}
\centering
\caption{Dissociation of charmonium in the magnetic field}
\label{table3}
\begin{tabular}{|p{0.7in}|p{1.4in}|p{1.3in}|p{1.2in}|} \hline
$c\overline{c}$ &T=2.5$T_{c}$ &T=2.495$T_{c}$ &T=2.490$T_{c}$ \\ \hline
${\alpha}=1$ &$eB={22.5m_{\pi}}^2$ &$eB={23m_{\pi}}^2$ &$eB={24{m}_{\pi}}^2$ \\ \hline
${\alpha}=0.5$ &$eB={20.0m_{\pi}}^2$ &$eB={21m_{\pi}}^2$ &$eB={22{m}_{\pi}}^2$ \\ \hline
\end{tabular}
\end{table}
By taking thermal medium at $T=2.5T_c$, the binding energy of charmonium dissociates while the magnetic field increases up to $eB={22.5m_{\pi}}^2$. By decreasing the temperature of the medium up to $T=2.490T_{c}$, the binding energy dissociated at $eB=24{m}_{\pi}^{2}$. A similar situation is also observed for the dissociation of bottomonium. However, the dissociation of bottomonium is larger than the dissociation of charmonium. This conclusion is in good agreement with following works in Ref:[13, 14,25].

\begin{table}
\centering
\caption{Dissociation of bottomonium in the magnetic field}
\label{table4}
\begin{tabular}{|p{0.7in}|p{1.4in}|p{1.3in}|p{1.2in}|} \hline
$b\overline{b}$ &T=2$T_{c}$ &T=1.99$T_{c}$ &T=1.98$T_{c}$ \\ \hline
${\alpha}=1$ &$eB={41m_{\pi}}^2$ &$eB=42{m_{\pi}}^2$ &$eB={43m_{\pi}}^2$ \\ \hline
${\alpha}=0.5$ &$eB={35m_{\pi}}^2$ &$eB=37{m_{\pi}}^2$ &$eB={38m_{\pi}}^2$ \\ \hline
\end{tabular}
\end{table}

\section{Conclusion}\label{5}
The SE is analytically solved by conformable fractional of the NU method, where the real fractional potential includes temperature $T$ and $eB$. The eigenvalues of energy and corresponding wave functions were obtained, in which they depend on the fractional parameter ${0}<{\alpha}\leq{1}$. The study showed the effect of fractional parameter on the effective interaction potential, the binding energy, dissociation of quarkonium in which the interaction potential is screened by increasing the fractional parameter. The binding energy and the dissociation of temperature in the fractional quark model were found to the lower than the classical quark model at ${\alpha}$=1. We have also observed that the magnetic field is largely affected by large-distance interaction, as a result of which the real part of potential is more attractive. The sound representation of the fraction solution provides an efficient and elegant way to solve the specific problems on the physics of interest. Consequently, studying of analytical solution of the modified fractional radial Schr\"{o}dinger equation  for the hot-magnetized interaction  potential within the framework conformable fractional the Nikiforov-Uvarov method could provide valuable information on the quantum mecahnical  dynamics at nuclear, atomic and molecule physics and opens new window.

\noindent {}

\noindent \textbf{References}

\noindent [1] R. Alexander, \textit{Phys. Reports}~\textbf{858}, 1-117 (2020).

\noindent [2] B. Jean-Paul and M.A. Escobedo, \textit{J. High Energy Phys.}~\textbf{6}, 1 (2018).

\noindent [3] M. Abu-Shady, H. M. Mansour, and A. I. Ahmadov, \textit{Adv.~High Energy Phys.} vol. \textbf{2019}, ID\textit{ }4785615 (2019).

\noindent [4] M. Abu-Shady and A. N. Ikot,~\textit{Eur. Phys. J. Plus}~\textbf{134}, 321 (2019).

\noindent [5] M. Abu-Shady, and M. Soleiman, \textit{Phys. Particles and Nuclei Lett.} \textbf{10}, 683 (2013).

\noindent [6] M. Abu-Shady, \textit{Mod. Phys. Lett. A} \textbf{29}, 1450176 (2014).

\noindent [7] M. Abu-Shady and H. M. Mansour, \textit{Phys. Rev. C}~\textbf{85}, 055204 (2012).

\noindent [8] V. Skokov, A. Illarionov, and V. Toneev, \textit{Int. J. Mod. Phys. A} \textbf{24}, 5925 (2009).

\noindent [9] V. Voronyuk, V.D. Toneev, W. Cassing, E.L. Bratkovskaya, V.P. Konchakovski, and S.A. Voloshin, \textit{Phys. Rev. C} \textbf{83}, 054911 (2011).

\noindent [10] C.S. Machado, F. S. Navarra, E. G. de Oliveria, and J. Noronha, \textit{Phys. Rev. D} \textbf{88}, 034009 (2013).

\noindent [11] M. Hasan, B. Chatterjee, B. K. Patra, \textit{Eur. Phys. J. C} \textbf{77}, 767 (2017).

\noindent [12] P. Bagchi, N. Dutta, B. Chatterjee, S. P. Adhya, arxiv: 1805.04082V1 (2018).

\noindent [13] C. Bonati, M. D'Elia, and A. Rucci, \textit{Phys. Rev. D} \textbf{92}, (2015).

\noindent [14] M. Abu-Shady, and H. M. Fath-Allah, arxiv2104.00545 (2021).

\noindent [15] A. Rothkopf, \textit{Phys. Rept.} \textbf{858}, 1(2020).

\noindent [16] M. Abu-Shady, and Sh. Y. Ezz-Alarab. \textit{Few-Body Systems}~\textbf{62.2}:1-8 (2021).

\noindent [17] Al-Jamel, \textit{Int. J. Mod. Phys. A} \textbf{34}, 1950054, (2019).

\noindent [18] T. Abdeljawad,\textit{J. Comput. Appl. Math}. \textbf{259}, 57, (2015).

\noindent [19] R. Herrmann, arXiv /0510099v4 (2006).

\noindent [20] M. Abu-Shady, \textit{Int. J. Mod. Phys. A} \textbf{34}.31: 1950201 (2019).

\noindent [21] I. Podlubny, Fractional Differential Equations (Academic Press, 1999).

\noindent [22] M. Caputo and M. Fabrizio, \textit{Prog. Fract. Differ. Appl.} \textbf{1} 73, (2015)

\noindent [23] R. Khalil, M. A. Horani, A. Yousef, and M. Sababheh, \textit{J. Comput. Appl. Math.} \textbf{264}, 65 (2014).

\noindent [24] H. Karayer, D. Demirhan, and F.
B\"uy\"ukk{\i}l{\i}\c{c}, \textit{Commun. Theor. Phys.} \textbf{66},
12, (2016).

\noindent [25] R. Kumar and F. Chand, \textit{Commun. Theor. Phys.} \textbf{59}, 528 (2013).

\noindent [26] M. Hasan, B. K. Patra, arxiv:200412857 V1 (2020).

\noindent [27] E. J. Ferrer, V. de la Incera and X. J. Wen, \textit{Phys. Rev. D} \textbf{91}, 054006 (2015).

\noindent [28] Yu. A. Simonov, \textit{Phys. At. Nucl.} \textbf{58}, 107 (1995).

\noindent [29] M. A. Andreichikov, V. D. Orlovsky and Yu. A. Simonov, \textit{Phys. Rev. Lett.} \textbf{110}, 162002 (2013).

\noindent [30] M. Hasan, B. K. Patra, B. Chatterjee, and P. Bagchi, arXiv:1802.06874 (2018).

\noindent [31] J. Beringer, Particle Data Group, Review of Particle Physics, \textit{Phys. Rev. D} \textbf{86}, 010001 (2012).

\noindent [32] V. K. Agotiya, V. Chandra, M. Y. Jamal, and I. Nilima, \textit{Phys. Rev. D} \textbf{94}, 094006 (2016).
\end{document}